\documentclass[11pt]{article}
\usepackage{amsmath}
\usepackage{amssymb}
\usepackage{authblk}

\title{An unexpected theoretical structure that could explain quantum-mechanics postulates like the Born rule and the wave-function reduction}
\author[1]{L\'{e}on Brenig}
\author[2]{Marc Vincke}
\affil[1]{Faculty of Sciences, Universit\'{e} Libre de Bruxelles, 1050 Brussels, Belgium - Corresponding author (leon.brenig@ulb.be)}
\affil[2]{Faculty of Applied Sciences, Universit\'{e} Libre de Bruxelles, 1050 Brussels, Belgium}

\begin{document}

\maketitle

\begin{abstract}
A unique postulate is shown to underly the whole quantum mechanics theory: the invariance of the Heisenberg uncertainty inequality under a group of special nonlinear gauge transformations (NLGT). With this postulate, the quantum mechanics of a free particle is derived from classical mechanics, including the statements of the postulates of quantum mechanics, except for the wave-function-collapse postulate. An explanatory mechanism for the latter postulate is derived by performing an analytical continuation of the NLGTs.  This extension results in a Schr\"{o}dinger-bridge process, intertwined under the NLGT with the standard unitary quantum evolution, and revealing non-quantum (or beyond-quantum) phenomena.  Mechanisms of that latter kind, like the ones associated to the quantum measurement process, occur in a new space-like dimension and hence are non causal in nature, in opposition to a time evolution.
The present exercice focusses on the free particle in order to highlight the features of the performed derivation in the simplest possible way. Work is in progress to extend the performed derivation beyond that simple case.
\end{abstract}

\section*{Keywords}
Quantum Mechanics,
Classical mechanics,
Schrödinger equation,
Stochastic bridge,
Wave-function collapse,
Born rule.

\section{Introduction}
In an article published in 1931 \cite{Schrodinger:1931:UNG} Erwin Schr\"{o}dinger concluded:
``But I do not wish to analyze these points more closely before
time tells if they can really lead to a better understanding of quantum
mechanics''. These were the last words of an article in which
Schr\"{o}dinger built a classical random process intended to understand
the apparition of the Born rule in quantum mechanics. This stochastic
problem was the following: a gas of independent classical diffusing
particles is considered. Its position probability density $\rho(x,t)$
is measured at two times $t_{0}$ and $t_{1}$. At $t_{0}$ the obtained
distribution is $\rho_{0}(x)$ while at $t_{1}$ it is $\rho_{1}(x)$.
The problem lies in the fact that $\rho_{1}(x)$ is different from the one expected
for a classical diffusion process by propagating the initial distribution
$\rho_{0}(x)$. This would be called nowadays a large deviation phenomenon.
Such a deviation is always possible, though, rather unlikely.

Schr\"{o}dinger found a brilliant solution to this question. The evolution
of the position probability density between the two times $t_{0}$
and $t_{1}$ is given in the form of a system of two evolution equations,
one forward in time for a real positive function $\varphi(x,t)$ and
a second backward in time for another real positive function $\phi(x,t)$.
He showed that the system needs two boundary conditions: an initial
function for the forward equation and a final function for
the backward equation. However, neither of these functions $\varphi$ and
$\phi$ are probability densities. Nevertheless, their product gives
the answer to the problem: the probability density that interpolates
for any time $t$ in the interval $[t_{0},t_{1}]$ between the initial 
and final probability densities, $\rho_{0}(x)$ and $\rho_{1}(x)$,
is given by $\rho(x,t)=\varphi(x,t)\phi(x,t)$.  This expression is a natural analog of the Born rule.

Schr\"{o}dinger showed that a couple of nonlinear integral equations allows
for determining the initial condition of the foreward equation, $\varphi(x,t_{0})$,
and the final condition of the backward equation, $\phi(x,t_{1})$
from the given boundary densities $\rho_{0}(x)$ and $\rho_{1}(x)$
of the problem. As can be seen from the above product formula for
the probability density $\rho(x,t)$, it bears a strong similarity
with the Born rule $\rho(x,t)=\psi(x,t)\psi^{*}(x,t)$ where $\psi$
is the wave function solving the Schr\"{o}dinger equation and $\psi^{*}$
is its complex conjugate. Moreover, the system of the forward and
the backward equations is identical to the system of equations obtained
by replacing the time $t$ by $it$ in the Schr\"{o}dinger equation and
its complex conjugate. 
For a recent overview and contextualization of Schr\"{o}dinger's 1931 article, see \cite{chetriteSchrodingers1931Paper2021}.
These were intriguing results, however, the sad conclusion of the 1931 article
did not push the physicists to investigate more along this line of
reasoning, until the 80's, when Zambrini introduced Euclidean Quantum Mechanics as the closest analogy with Quantum Mechanics \cite{PhysRevA.35.3631}. This was not the case of the mathematicians and engineers
as this Schr\"{o}dinger's article and a second one published in 1932 \cite{schrodinger_sur_1932}
paved the way to the Stochastic Optimal Control theory \cite{leonardSurveySchrodingerProblem2014} \cite{miangolarraInferringPotentialLandscapes2024} 
 that is flourishing nowadays. In these theories the problem raised by Schr\"{o}dinger
in 1931 is called the Schr\"{o}dinger Bridge process.

The present work shows that the 1931 Schr\"{o}dinger's attempt was not
in vain. We, indeed, find that under a group of special nonlinear
gauge transformations, the Schr\"{o}dinger Bridge equations intertwine
with the unitary quantum evolution equation, i.e., the standard Schr\"{o}dinger
equation. We show that the two kinds of evolutions, Schr\"{o}dinger Bridge
and unitary evolutions, emerge from classical mechanics via the requirement
that the Heisenberg uncertainty principle remains invariant under
the foregoing nonlinear gauge transformations. 

A possible physical interpretation of the Schr\"{o}dinger Bridge evolution
is that it represents the collapse of the wave function during a measurement.
The reasoning leading to this claim is that a measurement apparatus
is a device that is designed in order to produce a collapse of the
wave function on an eigenvector of the measured observable. So, the
measurement process is an evolution with a final constraint. Of course,
it also has an initial condition that is the wave function $\psi(x,t_{0})$
at the time $t_{0}$ at which the measurement begins. Thus, the quantum
measurement process is conditioned by an initial and a final condition.
As such, the time parameter that parametrizes the measurement evolution
has a somewhat space-like aspect. This is what is also observed in
the Schr\"{o}dinger Bridge evolution: the evolution parameter of that
process is space-like. Indeed this parameter, let us call it $\tau$,
form a couple $(t,\tau)$ with the time $t$ of the unitary evolution.
It is shown in the sequel that under the nonlinear gauge transformations
this couple undergoes a hyperbolic rotation in which $\tau$ plays
the role of a space-like coordinate, as opposed to the unitary time $t$.

The article is structured as follows. Chapter 2 describes the requirement
of the invariance of the Heisenberg uncertainty principle under a
group of special nonlinear gauge transformations and its consequences.
In Chapter 3, we show that the invariance of the Heisenberg principle
results in a modification of the classical average-energy function
that becomes identical to the quantum average-energy function. In
this Chapter is also introduced a functional Lie algebra framework
that is essential to the theory presented here. Chapter 4 shows that
the standard Schr\"{o}dinger equation is obtained from the foregoing invariance
of the Heisenberg principle. It is also shown that all the postulates
of quantum mechanics, including the Born rule, can be derived at once
from the postulate of invariance of the Heisenberg principle, except
that of the wave-function collapse. In Chapter 5 we explain how extending
the parameter of the nonlinear gauge transformation to the complex
numbers generates a new energy-like functional that, joined with the
quantum energy functional, undergoes a hyperbolic rotation during
a nonlinear gauge transformation. In Chapter 6 the emergence of the
Schr\"{o}dinger Bridge process is explained and the link with the Schr\"{o}dinger
1931 article is made. Chapter 7 is devoted to showing that the wave-function
collapse during a measurement can be described as a Schr\"{o}dinger Bridge process. In
Chapter 8 we derive the hyperbolic rotation undergone by the couple
$(t,\tau)$ during a nonlinear gauge transformation. This leads to
a discussion of the nature of the evolution parameter $\tau$ of the
Schr\"{o}dinger Bridge process as a new space-like dimension. Chapter
9 presents the conclusions and perspectives of the present work.
 
\section{Heisenberg-inequality invariance under NLGT}
In the classical limit ($\hbar\rightarrow0$), the Schr\"odinger equation turns into a specific form of classical mechanics, based on the Hamilton-Jacobi and continuity equations, describing the dynamics of a statistical ensemble of identical classical particles \cite{landauMecaniqueQuantique1967}. For a single free particle of mass $m$, the Hamilton-Jacobi equation reads
\begin{equation}
  \label{eq-H-J}
  \partial_t s = -\frac{|\nabla s|^2}{2m}
\end{equation}
and the continuity equation takes the form
\begin{equation}
  \label{eq-cont}
  \partial_t \rho = -\nabla\cdot (\rho\frac{\nabla s}{m}),
\end{equation}
where $\rho(x,t)$ is the probability density of the ensemble, depending on the space coordinates $x$ and the time $t$, and $s(x,t)$ is the momentum potential or action, related to the particle momentum with coordinates $p$ by
\begin{equation}
  \label{eq-p}
  p=\nabla s.
\end{equation}

In order to build the quantum mechanics of an individual spinless particle starting from equations (\ref{eq-H-J}, \ref{eq-cont}), the present work relies upon a single postulate, dispensing with the usual postulates of quantum mechanics \cite{cohen-tannoudjiMecaniqueQuantique1977}. It starts from remarking that in the classical limit ($\hbar\rightarrow0$), the lower bound of the Heisenberg uncertainty relation is $0$, as explained below.  The classical uncertainties involved here are given below  in the particle average rest frame with the average position at the origin. In this frame, called the rest frame in the sequel, these classical uncertainties have the following forms
\begin{equation}
  \label{eq-D2-x}
  D^2_x=\int d^3 x\,x^2 \rho(x)
\end{equation}
for the position and
\begin{equation}
  \label{eq-D2-p}
  D^2_p=\int d^3 x\,p^2 \rho(x)= \int d^3 x\,|\nabla s|^2 \rho(x)
\end{equation}
for the momentum, the latter being related to the particle  classical average energy
\begin{equation}
  \label{eq-avg-E-cl}
  \mathcal{H}_{\rm cl}=\frac{D^2_p}{2m}.
\end{equation}

The expressions (\ref{eq-D2-x}) and (\ref{eq-D2-p}) of the two uncertainties can vanish, for instance when the probability density $\rho$ of the particle in its rest frame is a Dirac delta, reflecting exact knowledge. Hence, these uncertainties depend on
the probability density $\rho$ in such a way that $D_{x}^{2}\geq0$,
$D_{p}^{2}\geq0$. The only general inequality that their product
satisfies is, therefore, $D_{x}^{2}D_{p}^{2}\geq0$. Clearly, this
product does not fulfill the Heisenberg uncertainty inequality, and
this is what is expected for a classical description. Quantum mechanics,
however, as we now show, naturally emerges in this context when requiring
the Heisenberg uncertainty inequality to be fundamental and demanding
its invariance under a group of special transformations. These transformations
are the nonlinear gauge transformations (NLGT) \cite{doebnerIntroducingNonlinearGauge1996}
\begin{equation}
  \label{eq-NLGT}
\rho\rightarrow\rho(\alpha)=\rho\:;\:s\rightarrow s(\alpha)=e^{-\alpha}s\:;\:\alpha\in\mathbb{R}.
\end{equation}
Obviously, they represent scalings of the action $s(x,t)$ and, as
such, play a role in singling out the special role of the quantum
action constant $\hbar$, as we now see.

In order to be fulfilled the Heisenberg principle requires modifying
the definitions of the two classical uncertainties $D_{x}^{2}$, $D_{p}^{2}$
as follows
\begin{align}
	\label{eq-sigma-x}
D_{x}^{2}\Rightarrow\sigma_{x}^{2}&=D_{x}^{2}, \\
	\label{eq-sigma-p}
D_{p}^{2}\Rightarrow\sigma_{p}^{2}&=D_{p}^{2}+\frac{\hbar^{2}}{4}\frac{1}{\Delta_{x}^{2}},
\end{align}
where $\Delta_{x}$ is the Fisher length \cite{coxTheoreticalStatistics1979} (i.e., the inverse
of the Fisher information), a measure of the position uncertainty
that is an alternative to $D_{x}$. It is defined by 
\begin{equation}
  \label{eq-Fisher}
\frac{1}{\Delta_{x}^{2}}=4\int d^{3}x\mid\nabla\rho^{1/2}\mid^{2}
\end{equation}
and satisfies the Cram\'{e}r-Rao inequality \cite{coxTheoreticalStatistics1979}, $\Delta_{x}\leq D_{x}$
. From these definitions, one immediately realises, first, that the
product $\sigma_{x}^{2}\sigma_{p}^{2}$ obeys the Heisenberg inequality
\begin{equation}
  \label{eq-Heisen0}
\sigma_{x}^{2}\sigma_{p}^{2}\geq\frac{\hbar^{2}}{4}
\end{equation}
and, second, that under the NLGT of parameter $\alpha$ this product
transforms as
\begin{equation}
  \label{eq-Heisen-demo}
\sigma_{x}^{2}(\alpha)\sigma_{p}^{2}(\alpha)=e^{-2\alpha}D_{x}^{2}D_{p}^{2}+\frac{\hbar^{2}}{4}\frac{D_{x}^{2}}{\Delta_{x}^{2}}.
\end{equation}

The latter equation combined with the above cited Cram\'{e}r-Rao inequality
implies that
\begin{equation}
  \label{eq-Heisen}
\sigma_{x}^{2}(\alpha)\sigma_{p}^{2}(\alpha)\geq\frac{\hbar^{2}}{4}
\end{equation}
 is true for any $\alpha\in\mathbb{R}$. In other words, the Heisenberg
inequality is invariant under the NLGTs.
It must be stressed that this invariance is a direct consequence of
the above modification of $D_{p}^{2}$ into $\sigma_{p}^{2}$ which
involves the Fisher length. One, thus, can conclude that considering
as a Postulate the Heisenberg principle and its invariance under the
NLGTs imposes the change of definition from $D_{p}^{2}$ into $\sigma_{p}^{2}$
as described above. We now show that this result directly leads to
the emergence of quantum mechanics. 

It should be noted that the same result had already been obtained
by one of us in a previous work \cite{Bre-07} assuming invariance with respect
to isotropic space dilatations instead of the NLGTs.

These above remarkable properties bear some similarities with special relativity. In analogy with the fact that the velocity of light constitutes an upper limit for the velocities of material bodies, the parameter $\hbar^2/4$ represents a universal lower limit for the product of uncertainties $\sigma_x^2\sigma_p^2$. This product plays a role similar to velocity in the Lorentz transformations. Latter in the paper, the analogy with special relativity will appear even more striking.

For later comparison, let us stress that the classical theory, represented by the Hamilton-Jacobi (\ref{eq-H-J}) and continuity (\ref{eq-cont}) equations, is kept invariant under the NLGT (\ref{eq-NLGT}) provided that the time parameter is also dilated as
\begin{equation}
\label{eq-t-dilat}
t(\alpha)=e^\alpha t.
\end{equation}

\section{Derivation of the quantum average-energy functional}
Applying the uncertainty correction (\ref{eq-sigma-p}) to the classical average energy (\ref{eq-avg-E-cl}) leads, with the use of the Fisher-length definition (\ref{eq-Fisher}), to the following expression for the quantum average-energy functional
\begin{equation}
  \label{eq-avg-E-qu}
  \mathcal{H} = \frac{\sigma_p^2}{2m}=\frac{1}{2m}\int d^3 x [\rho |\nabla s|^2 + \hbar^2 |\nabla \rho^{1/2}|^2],
\end{equation}
where the second term in the integral is recognized as the quantum or Bohm potential \cite{bohmSuggestedInterpretationQuantum1952}, proportional to the inverse squared Fisher length according to (\ref{eq-Fisher}).
Expression (\ref{eq-avg-E-qu}) becomes
\begin{equation}
  \label{eq-avg-E-obs}
  \mathcal{H} = \frac{1}{2m}\int d^3 x \psi^* (-i\hbar\nabla)^2 \psi,
\end{equation}
after introducing the complex function $\psi$, defined by the canonical transformation
\begin{align}
  \label{eq-psi-polar}
  \psi&=\rho^{1/2} e^{i s/\hbar}, \\
  \label{eq-psi*-polar}
  \psi^*&=\rho^{1/2} e^{-i s/\hbar}.
\end{align}
The use of the term ``canonical'' will soon become clear.
The complex function $\psi$ is square integrable because $\rho$ is a probability density.  The polar form (\ref{eq-psi-polar}) is involved when discussing the classical limit of the wave function of a quantum system \cite{landauMecaniqueQuantique1967}.

The quantity $\mathcal{H}$ is called \emph{quantum} average energy because (\ref{eq-avg-E-obs}) takes the form of a quantum average of an observable, the Hamiltonian operator
\begin{equation}
\label{eq-H}
H=\frac{(-i\hbar\nabla)^2}{2m},
\end{equation}
acting on the Hilbert space of the wave functions $\psi$.

Quantum average of observables like (\ref{eq-avg-E-obs}) form a set that can be structured as a Lie algebra $\mathbb{Q}$. This is a Lie sub-algebra of the general Lie algebra $\mathbb{G}$ associated to the set of all functionals of the $\rho$ and $s$ functions and of their space derivatives \cite{PhysRevD.28.1916}.  The classical average energy (\ref{eq-avg-E-cl}) and the inverse square Fisher length (\ref{eq-Fisher}) belong to $\mathbb{G}$ but not to $\mathbb{Q}$. The power of the formalism based on the Lie algebra $\mathbb{G}$ is that it describes both classical mechanics and quantum mechanics with the same tools. The functional Lie brackets of the Lie algebras are given by expressions involving functional derivatives ($\frac{\delta\mathcal{A}}{\delta\rho(x)}, \cdots$),
\begin{equation}
\label{eq-Lie-G}
\{\mathcal{A},\mathcal{B}\}= \int d^3 x\, [\frac{\delta\mathcal{A}}{\delta\rho(x)}\frac{\delta\mathcal{B}}{\delta s(x)}-\frac{\delta\mathcal{B}}{\delta\rho(x)}\frac{\delta\mathcal{A}}{\delta s(x)}]
\end{equation}
for any couple of functionals $\mathcal{A},\mathcal{B}$ belonging to $\mathbb{G}$. Moreover, since the two pairs of conjugate functions $\rho, s$ and $\psi, \psi^*$ are related by the canonical transformation (\ref{eq-psi-polar}, \ref{eq-psi*-polar}), one gets
\begin{equation}
\label{eq-Lie-Q}
\{\mathcal{A},\mathcal{B}\}= \frac{1}{i\hbar}\int d^3 x\, [\frac{\delta\mathcal{A}}{\delta\psi(x)}\frac{\delta\mathcal{B}}{\delta \psi^*(x)}-\frac{\delta\mathcal{B}}{\delta\psi(x)}\frac{\delta\mathcal{A}}{\delta \psi^*(x)}]
\end{equation}
for any couple of functionals $\mathcal{A},\mathcal{B}$ belonging to $\mathbb{G}$.  Obviously, the Lie sub-algebra $\mathbb{Q}$ with the functional Poisson bracket (\ref{eq-Lie-Q}) is isomorphic to the commutator Lie algebra of quantum mechanics.

Under the NLGT (\ref{eq-NLGT}), the quantum average energy functional (\ref{eq-avg-E-qu}) transforms as
\begin{equation}
  \label{eq-avg-E-qu-al}
  \mathcal{H}(\alpha) = \frac{\sigma_p^2}{2m}=\frac{1}{2m}\int d^3 x [e^{-2\alpha}\rho |\nabla s|^2 + \hbar^2 |\nabla \rho^{1/2}|^2],
\end{equation}
or, given (\ref{eq-avg-E-obs}), as
\begin{equation}
  \label{eq-avg-E-obs-al}
  \mathcal{H}(\alpha) = \frac{1}{2m}\int d^3 x \psi^*(\alpha) (-i\hbar\nabla)^2 \psi(\alpha),
\end{equation}
where, using (\ref{eq-NLGT}) and (\ref{eq-psi-polar}, \ref{eq-psi*-polar}), one has
\begin{align}
  \label{eq-psi-polar-al}
  \psi(\alpha)&=\rho^{1/2} e^{i s(\alpha)/\hbar}=\rho^{1/2} e^{i e^{-\alpha}s/\hbar}=(\psi)^\frac{1+e^{-\alpha}}{2}(\psi^*)^\frac{1-e^{-\alpha}}{2}, \\
  \label{eq-psi*-polar-al}
  \psi^*(\alpha)&=\rho^{1/2} e^{-i s(\alpha)/\hbar}=\rho^{1/2} e^{-i e^{-\alpha}s/\hbar}=(\psi^*)^\frac{1+e^{-\alpha}}{2}(\psi)^\frac{1-e^{-\alpha}}{2}.
\end{align}
The formulae (\ref{eq-psi-polar-al}, \ref{eq-psi*-polar-al}) explain the use of the term ``nonlinear'' in the expression ``nonlinear gauge transformations'' used when introducing (\ref{eq-NLGT}).
The functional $ \mathcal{H}(\alpha)$ remains a functional belonging to the $\mathbb{Q}$uantum Lie algebra $\mathbb{Q}$ as (\ref{eq-avg-E-obs-al}) is a quantum average of the observable (\ref{eq-H}) acting on the Hilbert space of the complex square integrable wave functions $\psi(\alpha)$. This space is equivalent to the Hilbert space of the wave functions $\psi$ as the $\psi(\alpha)$ can be expanded on the $\psi$ basis. This result is consistent with the Heisenberg-inequality invariance (\ref{eq-Heisen}) that is the starting postulate of this work. From (\ref{eq-psi-polar-al}, \ref{eq-psi*-polar-al}), it is clear that the Born rule
\begin{equation}
\label{eq-Born-psi-al}
\rho=\psi^*(\alpha)\psi(\alpha)
\end{equation}
is conserved under the NLGT (\ref{eq-NLGT}).
 
\section{Quantum-mechanics postulates that can be disregarded}
The statements of all postulates of quantum mechanics \cite{cohen-tannoudjiMecaniqueQuantique1977}, except for the wave-function-collapse one, result from the present derivation of (\ref{eq-avg-E-obs}) from classical mechanics, as reviewed below. 

In classical mechanics, the state of the system is represented by the density $\rho$ and by the momentum $p=\nabla s$, thus also by the action $s$, noting consequently that the classical physics does not change when an arbitrary constant is added to $s$.  This property is transferred to the wave function $\psi$ through the function change (\ref{eq-psi-polar}).  The arbitrary phase factor of the wave function reflects in quantum mechanics the arbitrary additive constant related to the action in classical mechanics. 

The measurable physical quantities relating to the free particle are the position and the momentum.  The momentum is described in (\ref{eq-avg-E-obs}) by the hermitian operator $-i\hbar\nabla$ corresponding to the momentum in position representation.  The application of the spectral decomposition \cite{richtmyerPrinciplesAdvancedMathematical1985} to this observable results in eigenfunctions and related eigenvalues corresponding to a basis change to the momentum representation. The position is obviously described by an eigenvalue $x$ of the position representation.

The Born rule is derived from classical mechanics as follows. In the $(\rho,s)$ description
of classical mechanics given by equations (\ref{eq-H-J}, \ref{eq-cont}), the probability
density $\rho$ reflects the lack of a complete information on the
initial position of the particle. Classical mechanics propagates in
time this lack of information and does not add any new uncertainty
to that lack.

However, as shown above, imposing the Heisenberg inequality and its
invariance under the NLGT introduces a new contribution in the uncertainty
of momentum. This, in turn, modifies the energy functional as in equation
(\ref{eq-avg-E-qu}). As a result, the equation (\ref{eq-H-J}) for the action is modified by
a new term.  Indeed, calculating the time derivative of $s$ via the Lie bracket $\partial_{t}s=\{s,\mathcal{H}\}$, with $\mathcal{H}$ given by (\ref{eq-avg-E-qu}), one gets 
\begin{equation}
\label{eq-H-J-mod}
\partial_{t}s=-\frac{\mid\nabla s\mid^{2}}{2m}+\frac{\hbar^{2}}{2m}\frac{\nabla^{2}\rho^{1/2}}{\rho^{1/2}},
\end{equation}
which is a well-known result \cite{bohmSuggestedInterpretationQuantum1952}.
Note that the continuity equation (\ref{eq-cont}) for the probability density
$\rho$ remains unchanged. 

The unmodified system of equations (\ref{eq-H-J}, \ref{eq-cont}) admits a solution in which
$\rho$ is initially a Dirac delta and remains a Dirac delta for all
later times. This is not the case for the new system constituted by
the above modified equation for $s$ and the continuity equation.
This shows that the supplementary term in the equation for $s$ introduces
new randomicity in the evolution of the system that adds to the initial
lack of knowledge of the position.

Now, as is well-known, the system of the above modified equation for
the action $s$ along with the continuity equation (\ref{eq-cont}) is equivalent
to the Schr\"{o}dinger equation for $\psi$ under the canonical transformation (\ref{eq-psi-polar}, \ref{eq-psi*-polar}).
The latter directly implies that $\rho=\psi\psi^{*}$. The Born rule
is, thus, established. 

The above derivation of the Born rule is made in position representation.  It is also valid in momentum representation, as it is possible to consider an initial classical-mechanics formulation based on a momentum probability density instead of a position one \cite{rashkovskiyHamiltonJacobiTheoryHamiltonian2020}.  Consequently, the derived quantum average energy would be given in momentum representation instead of (\ref{eq-avg-E-obs}).

As any element of $\mathbb{G}$, $\mathcal{H}$ is the infinitesimal generator of a one-parameter Lie group of transformations.  Noting the group parameter $t$, the finite transformations are generated by the differential form, for any $\mathcal{A}$ of $\mathbb{G}$,
\begin{equation}
\label{eq-tr-H}
\partial_t \mathcal{A}=\{\mathcal{A},\mathcal{H}\}.
\end{equation}
When the wave function $\psi$ and its complex conjugate are taken for $\mathcal{A}$,  (\ref{eq-tr-H}) takes the form of the Schr\"odinger equations,
  \begin{align}
\label{eq-Schr}
i\hbar\partial_t \psi(x,t)&=H \psi(x,t), \\
\label{eq-Schr-cc}
i\hbar\partial_t \psi^*(x,t)&=-H \psi^*(x,t),
\end{align}
governing the unitary evolution with time of the wave function and of its complex conjugate.

\section{Emergence of a new average-energy-like functional and analytical continuation of the NLGTs}
The derivation of an explanatory mechanism associated with the wave-function-collap\-se postulate goes beyond the features related to the other postulates of quantum mechanics.  As already discussed above, those other postulates  result from the postulated Heisenberg inequality (\ref{eq-Heisen}) under the NLGT (\ref{eq-NLGT}), where $\alpha$ is assumed real. To open the path to non-quantum (or beyond-quantum) system properties, an analytical continuation of the NLGTs is contemplated.

Before doing so, let us first consider the transformation of the average energy functional $\mathcal{H}$ under the NLGT with real parameter $\alpha$, given by (\ref{eq-avg-E-qu-al}).
This yields
\begin{equation}
\label{eq-H-NLGT}
\mathcal{H}(\alpha)=e^{-\alpha}[\cosh(\alpha)\mathcal{H}-\sinh(\alpha)\mathcal{K}],
\end{equation}
where the new functional $\mathcal{K}$ is defined by
\begin{equation}
  \label{eq-K}
  \mathcal{K} =\frac{1}{2m}\int d^3 x [\rho |\nabla s|^2 - \hbar^2 |\nabla \rho^{1/2}|^2].
\end{equation}
Before discussing the meaning of this new functional let us transform it under a NLGT. This gives
\begin{equation}
\label{eq-K-NLGT}
\mathcal{K}(\alpha)=e^{-\alpha}[-\sinh(\alpha)\mathcal{H}+\cosh(\alpha)\mathcal{K}].
\end{equation}
Now, we should underline the following properties.

First, notice that in the classical limit ($\hbar\rightarrow0$), the two transformations (\ref{eq-H-NLGT}) and (\ref{eq-K-NLGT}) become identical, as
\begin{equation}
\label{eq-Hcl-NLGT}
\mathcal{H}(\alpha), \mathcal{K}(\alpha) \rightarrow \mathcal{H}_{\rm cl}(\alpha)=e^{-2\alpha}\mathcal{H}_{\rm cl}.
\end{equation}

Second, considering the two transformations (\ref{eq-H-NLGT}) and (\ref{eq-K-NLGT}) together, it clearly appears that they represent a hyperbolic rotation of the couple $(\mathcal{H}, \mathcal{K})$ followed by a dilatation.

Third, the new functional $\mathcal{K}$ that emerges from transformation (\ref{eq-H-NLGT}) has the physical dimension of an energy.  Due to that and due to the classical limit common to the average-energy functional $\mathcal{H}$, we call it the average-energy-like functional.

What could be the theoretical meaning of the average-energy-like functional $\mathcal{K}$? An important property of that quantity appears when considering the transformation (\ref{eq-H-NLGT}) of $\mathcal{H}$ in the case of $\alpha$ being a pure imaginary number, more precisely $\alpha=-i\pi/2$. In that case one gets
\begin{equation}
  \label{eq-from-H-to-K}
  \mathcal{H}(\alpha=-i\pi/2)=-\mathcal{K}.
\end{equation}
This relation prompts us to focus on the analytic continuation of the NLGT (\ref{eq-NLGT}). This would amount to consider arbitrary values of $\alpha=\alpha_R+\alpha_I$. However, we have not found any physical meaning in assigning any value to $\alpha_I$, except for integer multiples of $\pi/2$. Consequently, the set of $\alpha$ values considered here on is 
\begin{equation}
  \label{eq-alpha-C}
  \alpha=\alpha_R + i k \pi/2,
\end{equation}
where $k\in\mathbb{Z}$. For $k$ even the first term in the r.h.s. of (\ref{eq-Heisen-demo}) is positive, thus the invariance of the Heisenberg principle is maintained for the NLGT with such values of the parameter $\alpha$. For $k$ odd and in particular for $k=-1$ as in the transformation (\ref{eq-from-H-to-K}) yielding $\mathcal{K}$, the first term in the r.h.s. of (\ref{eq-Heisen-demo}) is negative so that invariance of the Heisenberg principle is lost, hence the wording ``beyond-quantum'' proposed above.

Applying the NLGT with $\alpha_0=-i\pi/2$ to the canonical transformation (\ref{eq-psi-polar-al}, \ref{eq-psi*-polar-al}) provides another canonical transformation from $\rho, s$ to a pair of conjugate real positive functions $\varphi, \phi$,
\begin{align}
\label{eq-G2B-varphi}
    \varphi&\equiv \psi(\alpha_0)=\rho^{1/2} e^{-s/\hbar},   \\
\label{eq-G2B-phi}
    \phi&\equiv  \psi^*(\alpha_0)=\rho^{1/2} e^{+s/\hbar},
\end{align}
conserving the Born rule 
\begin{equation}
\label{eq-Born-phi}
\rho=\psi^*(\alpha_0)\psi(\alpha_0)=\varphi\phi.
\end{equation}
Applying (\ref{eq-G2B-varphi}, \ref{eq-G2B-phi}) to (\ref{eq-K}) results in the form
\begin{equation}
\label{eq-K-phi}
\mathcal{K}=-\int d^3 x \phi H \varphi.
\end{equation}
It belongs to the set of functionals
\begin{equation}
\label{eq-elem-B}
\mathcal{A}=\int d^3 x \phi A \varphi,
\end{equation}
where $A$ is an observable, acting here on a Hilbert space of real positive functions, as established in \cite{PhysRevA.35.3631}, and which can be structured as a Lie sub-algebra $\mathbb{B}$ of $\mathbb{G}$. The functionals (\ref{eq-elem-B}) are clearly not quantum averages (the Hilbert spaces on which the observable acts are distinct), so that they do not belong to $\mathbb{Q}$. The functional Lie bracket (\ref{eq-Lie-G}) of $\mathbb{G}$ can be expressed in terms of the new functions $\varphi$ and $\phi$ as follows, 
\begin{equation}
\label{eq-Lie-B}
\{\mathcal{A},\mathcal{B}\}= \frac{1}{\hbar}\int d^3 x\, [\frac{\delta\mathcal{A}}{\delta\varphi(x)}\frac{\delta\mathcal{B}}{\delta \phi(x)}-\frac{\delta\mathcal{B}}{\delta\varphi(x)}\frac{\delta\mathcal{A}}{\delta \phi(x)}]
\end{equation}
for any couple of functionals $\mathcal{A},\mathcal{B}$ belonging to $\mathbb{G}$. The set of functionals of the form (\ref{eq-elem-B}) is closed under this functional Lie bracket. This raises that set to the status of a Lie sub-algebra of $\mathbb{G}$.

Under the NLGT (\ref{eq-NLGT}), with $\alpha\in\mathbb{R}$, the transformed average-energy-like functional (\ref{eq-K-NLGT}) can be rewritten with the account of (\ref{eq-K-phi}), as
\begin{equation}
  \label{eq-avg-E-like-al}
  \mathcal{K}(\alpha) = -\int d^3 x \phi(\alpha) H \varphi(\alpha),
\end{equation}
where
\begin{align}
  \label{eq-varphi-polar-al}
  \varphi(\alpha)&=\rho^{1/2} e^{-s(\alpha)/\hbar}=\rho^{1/2} e^{-e^{-\alpha}s/\hbar}, \\
  \label{eq-phi-polar-al}
  \phi(\alpha)&=\rho^{1/2} e^{+s(\alpha)/\hbar}=\rho^{1/2} e^{+e^{-\alpha}s/\hbar}.
\end{align}
The functional $ \mathcal{K}(\alpha)$ remains a functional of the Lie sub-algebra $\mathbb{B}$ of $\mathbb{G}$ as (\ref{eq-avg-E-like-al}) keeps the form (\ref{eq-elem-B}) and as in (\ref{eq-avg-E-like-al}), $H$ acts on the Hilbert space of the real positive functions $\varphi(\alpha), \phi(\alpha)$. This space is equivalent to the Hilbert space of the real positive functions $\varphi,\phi$, as for instance the $\varphi(\alpha)$ can be expanded on the $\varphi$ basis. The derivation is similar to the quantum-average-energy one, which results in (\ref{eq-avg-E-obs-al}), and leads to the same important conclusion that from (\ref{eq-varphi-polar-al}, \ref{eq-phi-polar-al}), the Born rule
\begin{equation}
\label{eq-Born-phi-al}
\rho=\phi(\alpha)\varphi(\alpha)
\end{equation}
is conserved under the NLGT (\ref{eq-NLGT}).

It can be concluded that when $\alpha\in\mathbb{R}$, the transformed functionals $\mathcal{H}(\alpha)$ and $\mathcal{K}(\alpha)$ remain in their original Lie algebras, $\mathbb{Q}$ and $\mathbb{B}$  respectively.  
 
\section{Emergence of the Schr\"{o}dinger-bridge process}
As any element of $\mathbb{G}$, $\mathcal{K}$ is the infinitesimal generator of a one-parameter Lie group of transformations.  Denoting the group parameter by $\tau$, the resulting differential transformation has the general form, for any $\mathcal{A}$ of $\mathbb{G}$,
\begin{equation}
\label{eq-tr-K}
\partial_\tau \mathcal{A}=\{\mathcal{A},\mathcal{K}\}.
\end{equation}
In the classical limit ($\hbar\rightarrow0$), as the functionals $\mathcal{H}$ and $\mathcal{K}$ both tend towards the same classical average energy $\mathcal{H}_{\rm cl}$, the differential transformations (\ref{eq-tr-H}) and (\ref{eq-tr-K}) become a unique transformation, so that the parameters $t$ and $\tau$ merge together ($\tau=t$),
\begin{equation}
\label{eq-tr-Hcl}
\partial_t \mathcal{A}=\{\mathcal{A},\mathcal{H}_{\rm cl}\}.
\end{equation}

When the positive functions $\varphi$ and $\phi$ are taken for $\mathcal{A}$,  (\ref{eq-tr-K}) results in the $\tau$ evolutions of those functions,
\begin{align}
\label{eq-evol-varphi}
    \hbar\partial_\tau \varphi(x,\tau)&=-H \varphi(x,\tau),   \\
\label{eq-evol-phi}
     \hbar\partial_\tau \phi(x,\tau)&  =+H \phi(x,\tau).
\end{align}
Equation (\ref{eq-evol-varphi}) describes a forward evolution in the parameter $\tau$, while equation (\ref{eq-evol-phi}) governs the corresponding backward evolution. This is similar to the fact that the Schr\"{o}dinger equation may be considered as representing a forward evolution in time $t$ for the wave function, while the complex conjugate equation corresponds to a backward evolution.
The choice of a notation $\tau$ that differs from the time $t$ results from the fact that the parameters $t$ and $\tau$ are expected to be independent, like the related Lie groups of transformations, with infinitesimal generators that are functionals belonging to two distinct Lie sub-algebras $\mathbb{Q}$ and $\mathbb{B}$ of $\mathbb{G}$. 

The equations (\ref{eq-evol-varphi}, \ref{eq-evol-phi}) were established by Schr\"{o}dinger \cite{Schrodinger:1931:UNG}\cite{schrodinger_sur_1932} in 1931 for the free particle, when trying to find a classical random process with characteristics as close as possible to those of the Schr\"{o}dinger equation (\ref{eq-Schr}). As stated at the beginning of the Introduction, Schr\"{o}dinger himself did not pursue this line of thinking, concluding his second article on that subject \cite{schrodinger_sur_1932} by saying that he did not see a link between this classical random process and quantum mechanics.  Due to that, his attempt has been overlooked by the physicists for a long time.

Indeed, at first sight, equations (\ref{eq-evol-varphi}, \ref{eq-evol-phi}) are as different as possible from quantum mechanics, being similar to classical heat equations. They have, however, been the starting point of a new
field of stochastic processes theories, under various names such as Schr\"{o}dinger bridges (denomination used from here on, hence the symbol chosen for the Lie algebra $\mathbb{B}$), Stochastic Bridges, Reciprocal
Processes, Bernstein Bridges, or Bernstein Processes \cite{Bernstein-1932}\cite{miangolarraInferringPotentialLandscapes2024}\cite{PhysRevA.33.1532}\cite{PhysRevA.35.3631}. Such theories, used in physics, biology, or finance, unveil the stochastic dynamics of systems based on marginal observations at different (at least two) points in time.  Schr\"{o}dinger's attempt, forgotten by the theoretical physicists since 1931, was reconsidered in the 80's as the closest classical analogy with quantum mechanics, and developed under the names of stochastic variational dynamics \cite{PhysRevA.33.1532} or Euclidean quantum mechanics \cite{PhysRevA.35.3631}.  The latter reference remarks, as a peculiar aspect of Euclidean quantum mechanics, that an analytical continuation $\tau= it$ allows one to switch from the Schr\"{o}dinger equations (\ref{eq-Schr}, \ref{eq-Schr-cc}) to the Schr\"{o}dinger-bridge equations (\ref{eq-evol-varphi}, \ref{eq-evol-phi}).

Schr\"{o}dinger's approach in\cite{Schrodinger:1931:UNG}\cite{schrodinger_sur_1932} assumed evolutions in the usual time $t$. As in the present work, $\tau$ is assumed independent of $t$, notations are chosen accordingly. Schr\"{o}dinger showed that, in order to be
solved in the finite $\tau'$ interval $[0,\tau]$, the system (\ref{eq-evol-varphi}, \ref{eq-evol-phi}) requires an initial
condition $\varphi(x,0)$ for the forward equation (\ref{eq-evol-varphi}) and a final
condition $\phi(x,\tau)$ for the backward equation (\ref{eq-evol-phi}). These two conditions
are provided by solving the following system of nonlinear integral
equations
  \begin{align}
\label{eq-Schr-sys-0}
    \varphi(x,0) \int dy\, h(x,0,y,\tau) \phi(y,\tau) &= \rho(x,0),   \\
\label{eq-Schr-sys-T}
    \phi(y,\tau) \int dx\, h(x,0,y,\tau) \varphi(x,0)&= \rho(y,\tau), 
\end{align}
where $\rho(x,0)$ and $\rho(y,\tau)$ are the two given probability
densities of the problem and $h(x,0,y,\tau)$ is the free-diffusion or
Gaussian kernel. Once the above Schr\"{o}dinger integral system is solved
for the the two boundary conditions $\varphi(x,0)$ and $\phi(x,\tau)$,
the system of evolution equations (\ref{eq-evol-varphi}, \ref{eq-evol-phi}) can, in turn, be solved
for any $\tau'\in[0,\tau]$ as follows
  \begin{align}
\label{eq-evol-0}
    \phi(x,\tau')&= \int dy\, h(x,0,y,\tau-\tau') \phi(y,\tau),   \\
\label{eq-evol-T}
    \varphi(y,\tau') &= \int dx\, h(x,0,y,\tau') \varphi(x,0).
\end{align}
It should be stressed here that the evolution equations (\ref{eq-evol-varphi}, \ref{eq-evol-phi}) in
$\tau$ is a two-boundary-values problem, in contrast to the Schr\"{o}dinger
equation (\ref{eq-Schr}) in $t$ which is an initial-value problem (a one-boundary-value
problem). The two boundary conditions $\varphi(x,0)$ and $\phi(x,\tau)$
solving the above nonlinear system of integral equations depend on
the two data, $\rho(x,0)$ and $\rho(y,\tau)$, corresponding to the
Schr\"{o}dinger's thought experiment for the diffusing particles explained
in the Introduction. In his 1931 article, Schr\"{o}dinger shows that the
ultimate solution of his thought experiment problem, that is, the
probability density $\rho(x,\tau')$ for any $\tau'\in[0,\tau]$, is given
by the product 
\begin{equation}
\rho(x,\tau')=\varphi(x,\tau')\phi(x,\tau'),
\end{equation}
a product formula that can be brought back to the Born rule through
the canonical transformation (\ref{eq-G2B-varphi}, \ref{eq-G2B-phi}) itself related to the NLGT (\ref{eq-NLGT})
for purely imaginary parameter value $\alpha=-i\pi/2$.

In the more general case of a particle in an external scalar potential, existence and uniqueness for an arbitrary $\tau$ value has been demonstrated (see \cite{PhysRevA.33.1532} and references therein) for both the boundary functions $\varphi(x,0)$ and $\phi(y,\tau)$ and the derived solutions (\ref{eq-evol-0}, \ref{eq-evol-T}) of the evolution equations (\ref{eq-evol-varphi}, \ref{eq-evol-phi}), provided that the boundary densities have no zeros, which is the case for a free particle. It is also worth mentioning that the above construction is symmetric with respect to reversal of the $\tau'$ interval $[0,\tau]$ ($\tau'\rightarrow\tau-\tau'$).

The wave function $\psi$ introduced by the canonical transformation (\ref{eq-psi-polar}, \ref{eq-psi*-polar}) can also be taken for $\mathcal{A}$ in the differential transformation (\ref{eq-tr-K}), instead of the positive functions $\varphi$ and $\phi$, taking the form (\ref{eq-K}) of $\mathcal{K}$. This approach was adopted in a previous work \cite{Bre-07}. A nonlinear Schr\"{o}din\-ger equation is then obtained instead of the Schr\"{o}dinger-bridge equations (\ref{eq-evol-varphi}, \ref{eq-evol-phi}).  This nonlinear Schr\"{o}din\-ger equation is a member of the general class of nonlinear Schr\"{o}dinger equations generated by a larger group of nonlinear gauge transformations \cite{doebnerIntroducingNonlinearGauge1996} among which those playing a central role in the present work, the NLGT (\ref{eq-NLGT}), form a sub-group. However, it does not allow the surprising properties revealed by the Schr\"{o}dinger bridge to be directly highlighted.

\section{Wave-function collapses as particular Schr\"{o}din\-ger-bridge processes}
\label{sec-collapse}
The choice of the boundary probability densities in the r.h.s. of the Schr\"{o}\-din\-ger system (\ref{eq-Schr-sys-0}, \ref{eq-Schr-sys-T}) is governed by measurements that can be made on the free particle, whatever the probability, possibly quite low, of their outcome.  It is assumed that $\rho(x,\tau'=0)$ corresponds to the system in a quantum state described by the wave function $\psi(x,t'=0)$ before measurement, so that $\rho(x,\tau'=0)=|\psi(x,t'=0)|^2$ according to the Born rule. 

A first example of measurement may consist of recording a boundary density $\rho(x,\tau)$ corresponding to the system in a quantum state described by the wave function $\psi(x,t'=t)$ that has evolved from $t'=0$ to $t'=t$ according to the Schr\"odinger equation (\ref{eq-Schr}). The boundary density is related to the wave function by the Born rule $\rho(x,\tau)=|\psi(x,t)|^2$, noting that the parameters $t$ and $\tau$ are independent ($t$ is given while $\tau$ is arbitrary).  This is illustrated by simple calculations provided in appendix and involving a Gaussian wave packet. 

As a second example, the boundary density $\rho(x,\tau)$ may correspond to the system in a quantum state immediately after a measurement of a physical quantity.  Taking the example of a particle position measurement providing the value $x_M$,  the resulting state can be described by the position eigenfunction with $x_M$ as corresponding eigenvalue (delta function centered on $x_M$). The density predicted by the Schr\"{o}dinger system inside the $\tau'$ interval behaves smoothly between its boundary values, as evidenced by simple calculations involving a narrowing and decentring Gaussian wave packet (see appendix). This example illustrates how the  Schr\"{o}dinger-bridge process provides a mechanism for the wave-function collapse, that the unitary evolution governed by the Schr\"odinger equation cannot accommodate. Note that the existence and the form of the wave-function collapse in the quantum measurement is still controversial and several theories are proposed to describe it, or even to rule it out \cite{giacosaUnitaryEvolutionCollapse2014} \cite{zurekDecoherenceEinselectionQuantum2003} \cite{burgosMeasurementProblemQuantum2015} \cite{ballentineFailureTheoriesState1991} \cite{masanesMeasurementPostulatesQuantum2019} \cite{brackenEnvarianceProbabilitiesQuantum2020}.  The difference of the present proposition with other theories of collapse is that the latter generally assume some modifications of the Schr\"{o}dinger equation. In this work, we do not suppose any modification of that equation. We derive from a single invariance principle two different kinds of dynamics, the unitary dynamics governed by the Schr\"{o}dinger equation and the dynamics corresponding to the Schr\"{o}dinger-bridge equations. The latter corresponds to the reduction process of the wave function.

An analogy can be made with the two-fences experiment, for which the boundary density $\rho(x,\tau'=0)$ corresponds to the state of the particles at the source side, and the boundary density $\rho(x,\tau'=\tau)$ corresponds either to the diffraction scheme appearing on the observation screen after emission of many particles from the source without impact-position measurement made on the observation screen in one case, or to the measured impact position on the screen for an individual emitted particle in the other case.

\section{Additional space-like dimension associated to the Schr\"{o}dinger-bridge process}
It is important to first address the case where the value $\alpha=i\pi$ is taken in the canonical transformation (\ref{eq-psi-polar-al}, \ref{eq-psi*-polar-al}) (thus taking $\alpha_R=0$ and $k=2$ in (\ref{eq-alpha-C})).  Then, the resulting canonical variables correspond to switching $\psi$ and $\psi^*$, so that the Schr\"{o}dinger equations (\ref{eq-Schr}, \ref{eq-Schr-cc}) are retrieved when time is reversed, $t\rightarrow -t$. It can thus be concluded that the NLGT (\ref{eq-NLGT}) extended to $\alpha=i\pi$ is a time-reversal operator.  Applying now the same NLGT to the canonical transformation (\ref{eq-G2B-varphi}, \ref{eq-G2B-phi}), results in those canonical transformations with $\alpha=+i\pi/2=-\alpha_0$ (thus taking $\alpha_R=0$ and $k=+1$ in (\ref{eq-alpha-C})) instead of $+\alpha_0$. The resulting canonical variables correspond to switching $\varphi$ and $\phi$, so that the Schr\"{o}dinger-bridge equations (\ref{eq-evol-varphi}, \ref{eq-evol-phi}) are retrieved when $\tau$ is reversed, $\tau\rightarrow -\tau$. It can thus be concluded that the NLGT (\ref{eq-NLGT}) extended to $\alpha=i\pi$ is also a $\tau$-reversal operator. 

The reversal operator has no impact here because the dynamics of the free particle are invariant to time and $\tau$ reversals.  This is not true for systems whose dynamics are not invariant to reversal.  Such systems, like a particle in a magnetic field, are out of the present scope. As a result, from now on, only the NLGT (\ref{eq-NLGT}) with $\alpha\in\mathbb{R}$ has to be addressed, as $\alpha=-i\pi/2$ is implied in the definition of $\mathcal{K}$ and as $\alpha=+i\pi/2$ and $\alpha=i\pi$ have no impact as a result of the invariance with time and $\tau$ reversal of the free-particle dynamics.

The transformations (\ref{eq-H-NLGT}, \ref{eq-K-NLGT}) of the average-energy and average-energy-like functionals $\mathcal{H}, \mathcal{K}$ show that the standard quantum unitary evolution and the Schr\"{o}dinger-bridge process are intertwined under the NLGT (\ref{eq-NLGT}).
The resulting functionals $\mathcal{H}(\alpha)$, $\mathcal{K}(\alpha)$ are themselves the infinitesimal generators of two independent one-parameter Lie groups, with parameters $t(\alpha)$ and $\tau(\alpha)$, so that, given (\ref{eq-H-NLGT}, \ref{eq-K-NLGT}) and (\ref{eq-tr-H}, \ref{eq-tr-K}), for any functional $\mathcal{A}$,
\begin{align}
\label{eq-tr-H-al}
    \partial_{t(\alpha)}\mathcal{A}&=e^{+\alpha}\{\mathcal{A},\mathcal{H}(\alpha)\}=[\cosh(\alpha)\partial_t-\sinh(\alpha)\partial_\tau] \mathcal{A},  \\
\label{eq-tr-K-al}
    \partial_{\tau(\alpha)}\mathcal{A}&=e^{+\alpha}\{\mathcal{A},\mathcal{K}(\alpha)\}=[-\sinh(\alpha)\partial_t+\cosh(\alpha)\partial_\tau] \mathcal{A}.
\end{align}
The multiplicative factor $e^{+\alpha}$ appearing
in front of the Lie brackets in (\ref{eq-tr-H-al}) and (\ref{eq-tr-K-al}) is due to the fact that
the NLGT (\ref{eq-NLGT}) is canonical only up to that factor. Indeed, the true
canonical transformation is $\rho\rightarrow\rho(\alpha)=e^{+\alpha}\rho\:;\:s\rightarrow s(\alpha)=e^{-\alpha}s$.
In order to preserve the normalisation of $\rho$ to $1$ in the transformation,
$\rho$ is replaced by $\frac{\rho}{\int d^{3}x\rho}$ everywhere
in the calculations of the Lie brackets. This results in the appearance
of the multiplicative factor $e^{+\alpha}$ in equations (\ref{eq-tr-H-al}, \ref{eq-tr-K-al}).
As a reminder, the Born rule is preserved by the NLGT itself: $\rho=\psi(\alpha)\psi^{*}(\alpha)=\varphi(\alpha)\phi(\alpha)$.

The transformed functionals $\mathcal{H}(\alpha)$, $\mathcal{K}(\alpha)$ take the forms (\ref{eq-avg-E-obs-al}, \ref{eq-avg-E-like-al}), so that the evolutions (\ref{eq-tr-H-al}, \ref{eq-tr-K-al}) result in particular in the unitary Schr\"{o}dinger equations (\ref{eq-Schr}, \ref{eq-Schr-cc}) and to the Schr\"{o}dinger-bridge equations (\ref{eq-evol-varphi}, \ref{eq-evol-phi}) respectively, with $\alpha$ put in index to the functions and the evolution parameters. It can be concluded that the system of equations made of the unitary Schr\"{o}dinger equations (\ref{eq-Schr}, \ref{eq-Schr-cc}) and of the Schr\"{o}dinger-bridge equations (\ref{eq-evol-varphi}, \ref{eq-evol-phi}) is invariant under the NLGT (\ref{eq-NLGT}).

The inversion of (\ref{eq-tr-H-al}, \ref{eq-tr-K-al}) provides the transpose of the Jacobian matrix $\frac{\partial(t(\alpha),\tau(\alpha))}{\partial(t,\tau)}$, resulting in the parameter transformation taking the form of a hyperbolic rotation,
\begin{align}
\label{eq-t-NLGT}
    t(\alpha)&=\cosh(\alpha)t+\sinh(\alpha)\tau,   \\
\label{eq-tau-NLGT}
    \tau(\alpha)&=\sinh(\alpha)t+\cosh(\alpha)\tau.  
\end{align}
The latter transformation displays a clear analogy with the Lorentz transformations in special relativity between the time $t$ and a spatial coordinate.  This analogy of the Schr\"{o}dinger-bridge parameter $\tau$ with a spatial coordinate is coherent with the consideration of boundary conditions instead of initial conditions when solving the Schr\"{o}dinger-bridge equations (\ref{eq-evol-varphi}, \ref{eq-evol-phi}).  Initial conditions are relevant for a causal process like the wave-function time evolution governed by the Schr\"odinger equation (\ref{eq-Schr}). Boundary conditions are relevant for a non-causal process like the one governed by the Schr\"{o}dinger-bridge equations (\ref{eq-evol-varphi}, \ref{eq-evol-phi}), for which a space-like parameter, $\tau$, is appropriate. This sheds light on the nature of the quantum measurement, which appears to be a non-causal process occurring in a new dimension that differs from time, between the states of the system before and just after the measurement of a physical quantity, both of them represented by histograms of data representing probability densities. After all, the device set
up to perform a given measurement on a physical system is designed
in order to provide a unique answer among an assigned set of possible answers.
As such, it necessarily contains a non-causal element. Remark, however, that the assumption of the existence of non-causal effects in the process of quantum measurement is not new and is a hotly debated subject \cite{boppTimeSymmetricQuantum2017}.

In the classical limit ($\hbar\rightarrow0$), the evolutions (\ref{eq-tr-H-al}, \ref{eq-tr-K-al}) as well as the parameters $t$ and $\tau$ merge together, and the couple of parameter transformations (\ref{eq-t-NLGT}, \ref{eq-tau-NLGT}) yields the time dilatation (\ref{eq-t-dilat}) that ensures the invariance of the Hamilton-Jacobi (\ref{eq-H-J}) and continuity (\ref{eq-cont}) equations under the NLGT (\ref{eq-NLGT}).

When investigating the nature of $\tau$ further on, as already noticed in \cite{Bre-07}, it can be proven that the sign of the derivative in $\tau$ of the squared Fisher length $\Delta_x^2$ (\ref{eq-Fisher}) is always the opposite of that of the quantum momentum uncertainty $\sigma_p^2$ (\ref{eq-sigma-p}),
\begin{equation}
\label{eq-der-tau-prod}
\partial_\tau \Delta_x^2 \ \partial_\tau \sigma_p^2 < 0,
\end{equation}
noting that in the classical limit $\hbar\rightarrow0$, the l.h.s.\ of (\ref{eq-der-tau-prod}) tends to zero.  The proof of (\ref{eq-der-tau-prod}) relies on the use of expressions (\ref{eq-avg-E-qu}), (\ref{eq-K}) and (\ref{eq-tr-K}). In other words, the uncertainties on both the conjugate observables, position and momentum, evolve in the opposite way with $\tau$. In a measurement of position one expects that the uncertainty on position would decrease while that on the momentum would grow. The opposite would happen in the case of a momentum measurement. This corresponds well with the inequality (\ref{eq-der-tau-prod}). This fact supports the hypothesis that the dynamics in $\tau$ corresponds to the collapse of the wave function. In contrast, the analog of expression (\ref{eq-der-tau-prod}) for the unitary evolution in time parameter $t$ reads
\begin{equation}
\label{eq-der-t-prod}
\partial_t \Delta_x^2 \ \partial_t \sigma_p^2 =0,
\end{equation}
as $\sigma_p^2$, proportional to the quantum average-energy functional $\mathcal{H}$ (\ref{eq-avg-E-qu}), is a conserved quantity in time $t$.

Another remarkable property, illustrating the nature of the two parameters $t$ and $\tau$, concerns  their cross derivative acting on any functional $\mathcal{A}$  belonging to the Lie algebra $\mathbb{G}$, which can be written under the form
\begin{equation}
\label{eq-cross}
(\partial_t \partial_\tau - \partial_\tau \partial_t) \mathcal{A}=[\partial_t, \partial_\tau] \mathcal{A} = R_{t\tau} \mathcal{A}.
\end{equation}
Moreover, using equations (\ref{eq-tr-H}) and (\ref{eq-tr-K}), along with the Jacobi identity,
one gets
\begin{equation}
\label{eq-cross-bis}
[\partial_{t},\partial_{\tau}]\mathcal{A}=\{\{\mathcal{H},\mathcal{K}\},\mathcal{A}\}=R_{t\tau}\mathcal{A}
\end{equation}
In (\ref{eq-cross}) and (\ref{eq-cross-bis}), $R_{t\tau}$ is a curvature operator in the sense defined in \cite{kauffmanNoncommutativityDiscretePhysics1998}, reflecting a possible non-commutativity of the cross derivatives.  
The situation is analogous with general relativity, where the Riemann curvature tensor, which is the commutator of two covariant-derivative components acting on a vector field (Ricci identity), is a direct measure of the curvature of a Riemannian manifold. In (\ref{eq-cross-bis}), the Lie
bracket $\{\mathcal{H},\mathcal{K}\}$ is directly related to such a
curvature.

Taking as functional $\mathcal{A}$ in (\ref{eq-cross}) a measure of the particle position uncertainty, namely the squared Fisher length $\Delta_x^2$ (\ref{eq-Fisher}), and using expressions (\ref{eq-avg-E-qu}), (\ref{eq-K}), (\ref{eq-tr-H}) and (\ref{eq-tr-K}), the resulting cross derivatives yield
\begin{equation}
\label{eq-cross-Fisher}
R_{t\tau} \Delta_x^2 = \frac{2\hbar^2}{m^2}\frac{1}{\Delta_x^2}= \frac{8\hbar^2}{m^2}\int d^3 x\, |\nabla\rho^{1/2}|^2.
\end{equation}
In other words, the action of the curvature operator on the squared Fisher length is proportional to the extra-term in the quantum average-energy functional (\ref{eq-avg-E-qu}), the so-called Bohm potential that distinguishes quantum mechanics from classical mechanics, bearing so  an analogy with general relativity. Note that in the classical limit $\hbar\rightarrow0$, the curvature tends to zero.

\section{Conclusion and perspectives}
A unique postulate is shown to underly the whole quantum mechanics theory: the invariance of the Heisenberg uncertainty inequality under a group of special nonlinear gauge transformations (NLGT). With this postulate, the quantum mechanics of a free particle is derived from classical mechanics, including the statements of the postulates of quantum mechanics, except for the wave-function-collapse postulate. An explanatory mechanism for the latter postulate is derived by performing an analytical continuation of the NLGTs.  This analytical continuation results in a Schr\"{o}dinger-bridge process, intertwined under the NLGT with the standard unitary quantum evolution, and revealing non-quantum (or beyond-quantum) phenomena.  Mechanisms of that latter kind, like the ones associated to the quantum measurement process, occur in a new space-like dimension and hence are non causal in nature, in opposition to a time evolution.

The new space-like dimension comes to light in an analogy with special relativity, when considering the Heisenberg uncertainty principle.
In this inequality the role of the Planck reduced constant $\hbar$
is that of a universal lower bound in the product of the uncertainties
on conjugated quantities like position and momentum. This is not without
analogy with the role of the light velocity $c$ in special relativity
as a universal upper limit for the velocity of any material body.
In both the theories the requirement of a universal finite boundary
imposes modifications in the relations defining energy, thereby, leading
to a change in the laws of dynamics in both cases. The new dimension behaves as a space-like one in a transformation with time that takes the form of a Lorentz transformation, which reinforces the analogy with special relativity. Moreover, the cross derivatives versus both parameters allows defining a curvature tensor, in analogy with the Ricci identity of general relativity.

The present exercice focusses on the free particle in order to highlight the features of the performed derivation in the simplest possible way. Work is in progress to extend the performed derivation beyond the free particle, by considering first a single particle in a scalar potential and in an external electro-magnetic field.  A scalar potential implies observables additional to the position and the momentum, with spectra that can be discrete. A magnetic field breaks the system invariance to time reversal.  The relation of time reversal with the NLGT, with the new space-like dimension, with the properties of the Schr\"{o}dinger-bridge process and more generally with the quantum measurement process and with the present derivation of quantum mechanics from classical mechanics, has to be clarified.  The extension to several interacting particles is under progress and an extension to quantum field theory is considered.

\appendix

\section{Simple illustrative model based on a Gaussian wave packet}
In cartesian coordinates, the Hamiltonian operator (\ref{eq-H}) is the sum of three terms that commute with each other, reflecting that the motions along the three coordinate axes are independent. Hence, the motion along a single space coordinate $x$ is addressed here.  The Schr\"{o}dinger equation reads
\begin{equation}
\label{eq-Schr-x}
i\partial_t\psi=-\frac{1}{2}\partial^2_{xx}\psi,
\end{equation}
after a choice of units based on an arbitrary angular frequency $\omega_0$ that fixes a time unit $T=(2\omega_0)^{-1}$, so that the length unit is $k_0^{-1}=\sqrt{\hbar T/m}$ and the energy unit is $2\hbar\omega_0=\hbar/T$.

The model relies on the choice of a simple initial wave packet depending on a width $\sigma>0$ as single parameter, with a wave function of the form (making the normalization factors implicit as they do not play any role in the rationales)
\begin{equation}
\label{eq-psi-x-0}
\psi(x,t'=0)=\exp(-\frac{x^2}{4\sigma}),
\end{equation}
corresponding to the initial uncertainties $\sigma_x^2(t'=0)=\sigma$ and $\sigma_p^2=2\mathcal{H}=1/4\sigma$ (without time dependence for the latter as the energy is a constant of motion). The resulting solution from $t'=0$ to $t'=t$ of the Schr\"{o}dinger equation (\ref{eq-Schr-x}) has the form
\begin{equation}
\label{eq-psi-x-t}
\psi(x,t)=\exp(-\frac{(1-i t/2\sigma)x^2}{4\sigma a(t)}),
\end{equation}
where the relative width $a(t)\geq1$ is given by
\begin{equation}
\label{eq-def-a}
a(t)=1+(t/2\sigma)^2.
\end{equation}
The resulting density and position uncertainty read
\begin{align}
\label{eq-rho-x-t}
 \rho(x,t)&=\exp(-\frac{x^2}{2\sigma a(t)}), \\
\label{eq-s2-x-t}
    \sigma_x^2(t)&=\sigma a(t).  
\end{align}

To illustrate the first example addressed in \S\ \ref{sec-collapse}, the unique Schr\"{o}dinger-bridge solutions (\ref{eq-evol-0}, \ref{eq-evol-T}) of the evolution equations (\ref{eq-evol-varphi}, \ref{eq-evol-phi}) are now being sought on a $\tau'$ interval $[0,\tau]$, with boundary conditions at $\tau'=0$ and at $\tau'=\tau$ corresponding to the evolution of the density (\ref{eq-rho-x-t}) according to the Schr\"{o}dinger equation (\ref{eq-Schr-x}), from $t'=0$ to $t'=t$. The derivation of those solutions requires the resolution of the Schr\"{o}dinger system (\ref{eq-Schr-sys-0}, \ref{eq-Schr-sys-T}). It takes into account the fact that the solutions are in the form of Gaussians, since the Gaussian kernel in (\ref{eq-Schr-sys-0}, \ref{eq-Schr-sys-T}, \ref{eq-evol-0}, \ref{eq-evol-T}) reads
\begin{equation}
\label{eq-kernel}
h(x,s,y,t)= [2\pi(t-s)]^{-1/2} \exp[-\frac{(y-x)^2}{2(t-s)}].
\end{equation}
Taking account of the Born rule (\ref{eq-Born-phi}), the resulting density evolution in the $\tau'$ interval reads
\begin{equation}
\label{eq-rho-tp-A}
\rho(x,\tau')=\exp(-\frac{x^2}{2\sigma A(\tau', \alpha)}),
\end{equation}
introducing the reduced width $A(\tau',\alpha)$, related to the position uncertainty as
\begin{equation}
\label{eq-def-A}
\sigma_x^2(\tau')=\sigma A(\tau',\alpha)=\sigma (1+\frac{1}{\alpha}\frac{\tau'}{\sigma})(1-(1-\frac{1}{\alpha})\frac{\tau'}{\sigma}).
\end{equation}
Taking now the boundary condition of (\ref{eq-rho-tp-A}) at $\tau'=\tau$ as the the density (\ref{eq-rho-x-t}) at the time $t$, according to (\ref{eq-Schr-sys-T}), provides the identity
\begin{equation}
\label{eq-identity-widths}
A(\tau,\alpha)=a(t).
\end{equation}
When solving (\ref{eq-identity-widths}), what is assumed to be given is a density (\ref{eq-rho-x-t}) at a given time $t>0$ resulting from the evolution governed by the Schr\"{o}dinger equation (\ref{eq-Schr-x}) from the initial condition (\ref{eq-psi-x-0}). Then, the resulting Schr\"{o}dinger-bridge solutions are reflected in a density behavior (\ref{eq-rho-tp-A}), where the $\alpha$ parameter of the reduced width (\ref{eq-def-A}) is obtained by solving (\ref{eq-identity-widths}).  The resulting $\alpha$ value depends by construction on $t$, assumed to be given, but also on $\tau$, which can be arbitrarily chosen.  It can be concluded that there is a continuous family of unique Schr\"{o}dinger-bridge solutions corresponding to a given Schr\"{o}dinger-equation solution with fixed initial condition and evolution time, having $\tau>0$ as parameter.

To illustrate the second example addressed in \S\ \ref{sec-collapse}, other unique Schr\"{o}\-din\-ger-bridge solutions (\ref{eq-evol-0}, \ref{eq-evol-T}) of the evolution equations (\ref{eq-evol-varphi}, \ref{eq-evol-phi}) on a $\tau'$ interval $[0,\tau]$, quite different of (\ref{eq-rho-tp-A}), can be sought by keeping the same boundary condition at $\tau'=0$ and by imposing as boundary condition at $\tau'=\tau$ a density corresponding to an accurate position measurement at $x=x_M$, 
\begin{equation}
\label{eq-density-x-pos}
\rho(x,\tau)=\exp(-\frac{(x-x_M)^2}{2\sigma B}),
\end{equation}
where the reduced width $B\rightarrow0$, as does the position uncertainty.  Taking account of the Born rule (\ref{eq-Born-phi}), the resulting density and position-uncertainty dependences on $\tau'$ read
\begin{align}
\label{eq-density-x-pos-tp}
\rho(x,\tau')&=\exp(-\frac{(x-x_M\tau'/\tau)^2}{2\sigma B(\tau')}), \\
\label{eq-unc-x-pos-tp}
\sigma_x^2(\tau')&= \sigma B(\tau').
\end{align}
In (\ref{eq-density-x-pos-tp}) and (\ref{eq-unc-x-pos-tp}), the reduced-width dependence has the form
\begin{equation}
\label{eq-B-tp}
B(\tau')=[1+(\frac{1}{\sigma}-1)\frac{\tau'}{\tau}](1-\frac{\tau'}{\tau}),
\end{equation}
and is equal to 1 for $\tau'=0$ and to 0 for $\tau'=\tau$. It is observed that on the $\tau'$ interval, the density centre evolves linearly from $0$ to $x_M$ and that the position uncertainty evolves quadratically from $\sigma$ to $0$ . This simple model provides an explanatory mechanism  for the wave-function collapse occurring when a physical quantity is measured, the particle position in the present case. The collapse occurs in the $\tau$ dimension, the time $t$ does not play any role.

\section*{Statements and Declarations}
The authors did not receive support from any organization for the submitted work.
The authors have no relevant financial or non-financial interests to disclose.

 \section*{Acknowledgements}
 It is a pleasure to thank Prof.\ Jean-Marc Sparenberg for encouraging us and organizing a seminar where 
 we exposed and discussed many advances presented in this article. We also would like to thank our 
 respective departments
at ULB University (Universit\'{e} Libre de Bruxelles), 
 Quantum Physics Research Unit (M.V.) and Dynamical Systems Physics (L.B.).

\bibliographystyle{plain}
\bibliography{mainbib.bib}

\end{document}